\documentclass[%
  aps,          
  preprintnumbers,
  prl,          
  twocolumn,    
  showpacs,     
  superscriptaddress, 
  longbibliography,   
]{revtex4-2}

\usepackage{amsmath,amssymb,amsfonts,amsxtra,mathrsfs,graphics,graphicx,amsthm,epsfig,bm,longtable,color,tikz,mathtools,xfrac,footnote}
\usepackage{dsfont}
\usepackage{lipsum}
\usepackage{textgreek}
\usetikzlibrary{decorations.pathmorphing}
\usetikzlibrary{decorations.markings}
\usetikzlibrary{quotes,arrows.meta}
\usetikzlibrary{arrows,decorations.markings,calc,fadings,decorations.pathreplacing,patterns,decorations.pathmorphing,positioning}
\usepackage{tikz-cd}
\usepackage{xcolor}
\usepackage{hyperref}
\usepackage{orcidlink}
\definecolor{PRLblue}{RGB}{0,0,180}
\hypersetup{
  hypertexnames=false,
  colorlinks=true,
  linkcolor=PRLblue,
  citecolor=PRLblue,
  urlcolor=PRLblue,
  pdftitle={A New Entropy Puzzle in ABJM},
}

\usepackage{physics}
\usepackage{braket}
\usepackage{ulem}

\newcommand{\bea}{\begin{equation}\begin{aligned}}
\newcommand{\eea}{\end{aligned}\end{equation}}
\newcommand{\be}{\begin{equation}}
\newcommand{\ee}{\end{equation}}
\newcommand{\ba}{\begin{align}}
\newcommand{\ea}{\end{align}}

\newcommand{\ii}{\mathrm{i}}

\newcommand{\rd}{\mathrm{d}}

\newcommand{\wt}{\widetilde}

\def\U{\mathrm{U}}
\def\SO{\mathrm{SO}}

\def\SU{\mathrm{SU}}

\begin{document}

\title{A New Entropy Puzzle in ABJM}

\author{Alexandre Belin}
\email{alexandre.belin@unimib.it}
\affiliation{Dipartimento di Fisica, Universit\`a di Milano-Bicocca, I-20126 Milano, Italy}
\affiliation{INFN, sezione di Milano-Bicocca, I-20126 Milano, Italy}

\author{Seyed Morteza Hosseini}
\email{morteza.hosseini@qmul.ac.uk}
\affiliation{Centre for Theoretical Physics, Department of Physics and Astronomy, Queen Mary University of London, London E1 4NS, United Kingdom}
\affiliation{School of Mathematics and Maxwell Institute for Mathematical Sciences, University of Edinburgh, Edinburgh EH9 3FD, United Kingdom}

\author{Palash Singh}
\email{palash.singh@unimib.it}
\affiliation{Dipartimento di Fisica, Universit\`a di Milano-Bicocca, I-20126 Milano, Italy}
\affiliation{INFN, sezione di Milano-Bicocca, I-20126 Milano, Italy}

\author{Rita Vadala}
\email{ritalucia.vadala@unimib.it}
\affiliation{Dipartimento di Fisica, Universit\`a di Milano-Bicocca, I-20126 Milano, Italy}

\author{Alberto Zaffaroni}
\email{alberto.zaffaroni@mib.infn.it}
\affiliation{Dipartimento di Fisica, Universit\`a di Milano-Bicocca, I-20126 Milano, Italy}
\affiliation{INFN, sezione di Milano-Bicocca, I-20126 Milano, Italy}

\begin{abstract}
We point out a puzzle about BPS black holes in AdS$_4\times S^7/\mathbb{Z}_k$: the entropy grows when $k$ increases although the number of  gauge-invariant  operators in the dual conformal field theory that can contribute to the entropy grows as $k$ decreases, due to the presence of monopole operators. Analyzing the superconformal index of ABJM, we show that the same puzzle exists for finite $N$ in the large charge limit. We propose a resolution  based on an increasing number of trace relations as $k$ decreases, which dominates over the other effect. We argue that this mostly affects fortuitous operators and we provide an explicit example of a fortuitous operator that disappears at small $k$.
\end{abstract}


\maketitle

\textit{Introduction}--%
The ABJM theory (Aharony--Bergman--Jafferis--Maldacena)  is a well-studied 3d supersymmetric conformal field theory which serves as the canonical example of AdS$_4$/CFT$_3$ \cite{Aharony:2008ug}. It is somewhat richer than $\mathcal{N}=4$ SYM, possessing in fact two holographic limits: one in terms of type IIA string theory and the other in terms of M-theory. Moreover, instead of a continuous gauge coupling, ABJM possesses a discrete coupling set by the level $k$ of its Chern-Simons term. This implies that even quantities protected by supersymmetry can depend non-trivially on $k$. One of the biggest surprises of ABJM theory was the scaling of its number of degrees of freedom with the rank of the gauge group $N$. This is best parametrized by the $S^3$ free energy, which, at large $N$ and finite $k$, obeys 
\be\label{S3}
F_{S^3}=-\log Z_{S^3} \sim \sqrt{k}N^{3/2} \,.
\ee
This result, inferred first from holography, was originally surprising because the field content of ABJM is bifundamental matter and one would therefore expect the free energy to grow like $N^2$. The reduction resulting in a lower power of $N$ is usually attributed to strong coupling. For the $S^3$ partition function, this puzzle was resolved when the free energy was computed directly in the CFT using localization \cite{Drukker:2010nc}.

In this paper, we raise a new puzzle in ABJM, this time for the superconformal index which has a direct interpretation as counting BPS local operators of the CFT. Using holography, we can again infer the $N$ and $k$ scaling of the degeneracies $a_n$ of the superconformal index as a function of the charge $n$
\begin{equation}
\label{entropyintro}
S(n) =\log |a_n|
\sim \sqrt{k} N^{3/2} f\left(\frac{n}{N^{3/2}\sqrt{k}}\right) \,, \end{equation}
where $f$ is a function of order one of its argument.
 This is the Bekenstein--Hawking formula for BPS black holes in AdS$_4$, translated to CFT parameters.

 The puzzle we raise is that there are actually \textit{more} possibilities for building gauge-invariant operators in the CFT when $k$ is small, due to more monopole sectors contributing at fixed quantum numbers. We would therefore expect to find an increase in the degeneracies as $k$ is decreased. The large $N$ gravity prediction,  \eqref{entropyintro}, goes  in the opposite direction. In this paper we study the $k$ dependence of the superconformal index at small $N$ and large charge $n$, and show that the asymptotic growth of the degeneracies also decreases as $k$ decreases, reinforcing the puzzle.  We note that the naive intuition of having fewer BPS operators at stronger coupling is  compatible with the $k$-scaling of the index, so the puzzle raised by the extra monopole sectors is in some sense doubly surprising.

 We propose to confront this question through the lens of trace relations and the fortuity program. This program proposes to separate BPS operators into two classes: \textit{monotones}, states that exist for all $N$ and have a smooth large $N$ limit in one-to-one correspondence with multi-graviton states, and \textit{fortuitous states}, which exist only up to a  threshold in $N$, beyond which they cease to be BPS \cite{Chang:2022mjp,Choi:2023znd}.
  Fortuitous operators are conjectured to be dual to black hole states in the bulk, while monotones should correspond to gravitons. We analyze the $k$-dependence of BPS counts for both monotones and fortuitous states and we conjecture the following structure: There are more monotones at small $k$, compatible with the Kaluza--Klein reduction of supergravity on $S^7/\mathbb{Z}_k$ as well as the CFT expectation that there are more monopole sectors from which we can build operators. On the other hand, there are fewer fortuitous states at small $k$ compatible with the gravity scaling \eqref{entropyintro}. We attribute this to an increase in the possible trace relations as $k$ is decreased. So while there are in principle more ways to make BPS operators, there are also more ways to have trace relations and it is this second effect that dominates. We provide a concrete example of a BPS fortuitous cohomology that disappears due to new trace relations as $k$ decreases. We conjecture that these new trace relations dominate for large quantum numbers, and that at large $N$ they eventually lead to the $N^{3/2}\sqrt{k}$ scaling of the entropy.

\textit{The  entropy puzzle in ABJM}---%
We are interested in the ABJM  three-dimensional superconformal field theories,  with $\mathcal N=6$ supersymmetry and global symmetry $\SU(4)_R\times \U(1)_t$, realized as Chern--Simons matter theories with gauge group $\U(N)\times \U(N)$ and Chern--Simons levels $k$ and $-k$.  In a 3d $\mathcal N=2$ formalism, the matter content consists of two pairs of bifundamental chiral multiplets, $A_i$ and $B_i$ with $i=1,2$, transforming in the representations $({\bf \overline N},{\bf N})$ and $({\bf  N},{\bf \overline N})$ of the gauge group, respectively, and interacting with the superpotential $ W = \frac{2\pi}{k} \epsilon_{ij}\epsilon_{pq}{\rm Tr}(A_i B_p A_j B_q)$. In $\mathcal N=2$ notation, only the symmetry $\U(1)_R\times \SU(2)\times \SU(2) \times \U(1)_t$ is manifest, where the two $\SU(2)$ factors rotate the chiral fields $A_i$ and $B_i$, respectively, and $\U(1)_t$ is a topological symmetry. In the limit of large $N$ and finite $k$,  the ABJM theory is holographically dual to M-theory on AdS$_4\times S^7/\mathbb Z_k$.  In the 't Hooft limit, i.e. $N,k \rightarrow \infty$, with $\lambda=N/k$ fixed and large, the dual is described by type IIA string theory on AdS$_4\times \mathbb{CP}^3$.  

It is well known that for $k=1,2$ there is a non-perturbative  enhancement to $\mathcal N=8$ supersymmetry and $\SO(8)_R$ symmetry. This is due to the existence of monopole operators,  local disorder operators creating  Dirac monopole singularities for the gauge fields at a specific point. The monopole operators also deeply affect the spectrum of BPS states for generic $k$. For ABJM,  the monopole operators $T^{({\bf m},\wt{{\bf m}})}$ are labeled by integer magnetic fluxes  under the two $\U(N)$ gauge groups,
\begin{equation}
 {\bf m}=(m_1,m_2,\ldots , m_N)\, ,\quad  \wt{\bf m} =(\wt{m}_1, \wt{m}_2,\ldots , \wt{m}_N) \, ,
\end{equation}
defined up to the Weyl group that permutes their entries. 
The operator $T^{({\bf m},\wt{{\bf m}})}$ has charge $\sum_{i=1}^N m_i=\sum_{i=1}^N \wt{m}_i$ under the topological symmetry $\U(1)_t$ and acquires electric charges under the  gauge groups due to the Chern--Simons couplings. A monopole transforms in a representation of $\U(N)\times \U(N)$ with highest weight $(-k {\bf m},k \wt{{\bf m}})$
when the fluxes are ordered as $m_1\ge m_2\ge \ldots \ge m_N$ (and similarly for $\wt{m_i}$) \cite{Klebanov:2008vq}. In particular the $\U(1)\times \U(1)$ charge of the monopole is $(- k \sum_i m_i, k \sum_j\wt m_j)$. The quantum corrections to the gauge and flavor charges of the monopole operators  \cite{Borokhov:2002cg,Benini:2011cma}  vanish  if $\wt{{\bf m}}={\bf m}$, which is mostly the case that we will consider. Gauge-invariant  operators are obtained by dressing the bare monopole operator with  bifundamental matter fields with compensating gauge charges.  The lowest dimension BPS operators are obtained by dressing the monopole $\wt{{\bf m}}={\bf m}= (\pm1,0,\dots,0)$ with $k$ bifundamental scalars $A_i$ or $B_i$ and have R-charge and dimension $k/2$. In particular, monopole operators have large dimension and disappear from the spectrum in the limit $k\gg 1$ while they are as abundant as the other BPS operators for $k=1$ or $k=2$.

We see that more and more monopole sectors can contribute to the BPS spectrum as $k$ is decreased. On the other hand,  the entropy of the BPS black holes in AdS$_4\times S^7/\mathbb Z_k$
scales as in \eqref{entropyintro} and decreases when  $k$ is decreased. 
This expression has been confirmed by explicit computations from the topologically twisted index in \cite{Benini:2015eyy,Bobev:2022eus} for magnetically charged BPS black holes and from the superconformal index in the Cardy limit in \cite{Choi:2019zpz,Choi:2019dfu,Bobev:2022wem} for electrically charged, rotating BPS Kerr-Newman black holes.  The entropy puzzle extends outside the holographic regime and holds also in the Cardy limit at finite $N$. We will indeed show that, for $N$ finite and $n\gg 1$,   the degeneracy of states behaves as $S(n)=\log |a_n|\sim  \sqrt{| w(k)| n}$ where $|w(k)|$ is a  monotonically non-decreasing  function of $k$.

In contrast,  the scaling \eqref{S3}  of the $S^3$ free energy  agrees with the expectations. Indeed, we know from holography that, for a background AdS$_4\times M$, where $M$ is a seven-dimensional Sasaki-Einstein manifold, $F_{S^3}\sim 1/\sqrt{{\rm Vol} (M)}$  and the volume of the orbifold $S^7/\mathbb Z_k$ decreases as $1/k$. The argument extends to the $S^3$ free energy with arbitrary R-charge assignment $\Delta$, since we can relate it to the volume of a Sasakian deformation of $S^7/\mathbb Z_k$. 

\textit{The Cardy limit at finite $N$---}%
The spectrum of BPS operators is captured by the superconformal index \cite{Bhattacharya:2008zy}
\begin{equation}
 \mathcal{ I}(x,v_1,v_2,t)
 ={\rm Tr}\,(-1)^F x^\mathcal{ D+J}
 v_1^{J_1}v_2^{J_2}t^T ,
\end{equation}
where $\cal D$ is the dilatation operator, $\cal J$ the angular momentum, and $J_{1,2}$ and $T$ are the generators of the 
$\SU(2)\times \SU(2)\times \U(1)_t$ symmetry preserved by the choice of a supercharge $\cal Q$.
Only states annihilated by $\cal Q$ and $\cal Q^\dagger$ contribute to the trace and are in one-to-one correspondence with the cohomology of $\cal Q$.
 The index  reads \cite{Kim:2009wb}
\bea
\label{SCI}
\mathcal I
&=\frac{1}{(N!)^2}
\sum_{{\bf m},\wt{\bf m}}
\oint\prod_{i=1}^N
\frac{{\rm d}u_i}{2\pi\ii u_i}
\frac{{\rm d}\wt u_i}{2\pi\ii\wt u_i}
\,u_i^{-k m_i}\wt u_i^{k\wt m_i}
\\
&\times
\prod_{i<j}^N
Z_{\rm vec}\!\left(
\frac{u_i}{u_j},m_i-m_j
\right)
Z_{\rm vec}\!\left(
\frac{\wt u_i}{\wt u_j},
\wt m_i-\wt m_j
\right)
\\
&\times
\prod_{i,j=1}^N\prod_{a=1}^2
Z_{\rm ch}\bigg(
y_{a}\frac{\wt u_i}{u_j},
\wt m_i-m_j
\bigg)Z_{\rm ch}\bigg(
y_{a+2}\frac{u_i}{\wt u_j},
m_i-\wt m_j
\bigg)
,
\eea
where 
\bea
Z_{\rm vec}(u,m)
&= x^{-|m|}\prod_{\sigma=\pm 1}
\left(
1-(-1)^{m}x^{|m|} u^\sigma\right ),
\\
Z_{\rm ch}(u,m)&=
\displaystyle
\left(x^{1/2}u^{-1}\right)^{|m|/2}
\frac{( (-1)^m x^{3/2+|m|}u^{-1};x^2)_\infty}
{( (-1)^m x^{1/2+|m|}u;x^2)_\infty}
\eea
where $(a;q)_\infty=\prod_{n=0}^{\infty}(1-aq^n)$.  In \eqref{SCI}, we used the integration over the diagonal $\U(1)$ to set $\sum_{i=1}^N m_i=\sum_{i=1}^N\wt m_i$. Then the dependence   on the topological  fugacity $t$, $\prod_{i=1}^N t^{m_i}$,  has been re-absorbed in the fugacities $(y_1,y_2,y_3,y_4)=(t^{-1/k}v_1^{-1},v_1 t^{-1/k},v_2^{-1}t^{1/k},v_2 t^{1/k})$ of the four fields $(A_1,A_2,B_1,B_2)$ with a change of integration variables. 

\begin{figure}[t]
    \centering
    \includegraphics[width=\columnwidth]{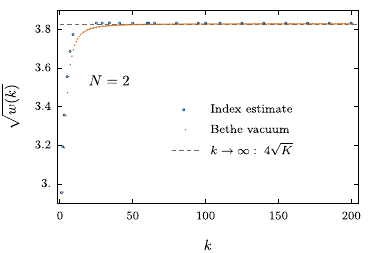}
    \caption{Comparison of $\sqrt{w(k)}$ for $N=2$.
Blue open circles: large-order envelope fits to the even-order superconformal-index coefficients. Orange dots: the selected Bethe-vacuum prediction at each integer $k=1,\ldots,200$. Dashed line: the
fixed-$N$, large-$k$ limit $\sqrt{w(\infty)}=4\sqrt{K}\approx 3.8282$.}
    \label{fig:num}
\end{figure}

For simplicity, we specialize to the case of a single fugacity $x=z^2$, $\mathcal{ I}(z)\equiv \mathcal{ I}(z^2,1,1,1)=\sum_{n=0}^{\infty}a_n z^n$, where we introduced $z$  in order to have a series with integer powers. 
  We performed a numerical analysis for $N=1,2$ and $k$ between $1$ and $200$,  expanding the index in powers of $z$ up to order $z^{1500}$, and using Richardson extrapolation. We obtain a series with  alternating integer coefficients that is compatible with the asymptotic behavior
  \begin{equation}\label{asympt} a_n \sim  e^{\sqrt{|w(k)|  n}}\cos(\phi_n(k)) n^{C(k)}\, .\end{equation}
  The degeneracy of states of charge $n$  is  given by $S(n)=\log |a_n|\sim\sqrt{|w(k)| n} $.                  For $N=1$, $w(k)= 4 K$ for all $k$, where $K=\beta(2)=0.91597\ldots$ is  Catalan's constant. For $N=2$, the function $\sqrt{w(k)}$ is given by the blue open circles in Figure \ref{fig:num} and is  monotonically increasing.  

\begin{figure}[t]
    \centering
    \includegraphics[width=\columnwidth]{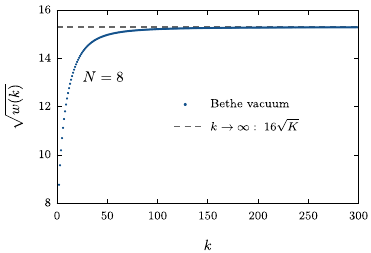}
    \caption{The quantity $\sqrt{w(k)}$ for $N=8$.
    Blue points: the twisted superpotential evaluated on the selected Bethe-vacuum branch for
    $k=2,\ldots,300$. This branch smoothly approaches the black hole
    saddle for $N\gg k$. Dashed line: the fixed-$N$,
    large-$k$ limit $\sqrt{w(\infty)}=16\sqrt{K}\approx 15.313$.}
    \label{fig:num2}
\end{figure}

The numerical result can be explained by the existence of a saddle point of the integral approaching a singularity $z_0$  on the unit circle 
\begin{equation}\label{saddlefin}
 a_n =\oint \frac{\rd z}{2\pi \ii } \frac{\mathcal{ I}(z)}{z^{n+1}}\sim \int \frac{\rd\epsilon}{z_0^n} \, \epsilon^\alpha e^{
-\frac{w}{ 2\epsilon}- \ii n  \epsilon} \sim \frac{e^{\sqrt{w n}(1+\ii)}}
{ z_0^{n}n^{\frac 34
+\frac{\alpha}{2}}}\,,
 \end{equation}
characterized by  $\log\mathcal{ I}\sim -\frac{w(k)}{2\epsilon}+\alpha(k)\log\epsilon $ for $z= z_0 e^{\ii\epsilon}\rightarrow z_0$, with $w(k)$ and $\alpha(k)$ real,  together with the existence of a complex conjugate saddle. 
For $N=1$ we can check this explicitly using
  \begin{equation}
  \mathcal{ I}_{N=1,k}(z)=\frac1k \sum_{l=0}^{k-1} \mathcal{ I}_\text{free}(z^2,1,1,e^{\frac{2\pi \ii l}{k}})\, ,    
  \end{equation}
  where $\mathcal{ I}_\text{free}$ is the index of four free chiral multiplets. For finite $k$,  the main contribution to the index comes from the two saddles $z=\pm \ii$ and the term $l=0$ ($\log \mathcal{ I}_\text{free}\sim \mp \frac{2K}{\epsilon}$). More generally, it is known that in the Cardy limit $z=e^{\ii\epsilon}\rightarrow 1$  the asymptotic behavior of the index is  $\log \mathcal{ I}
\sim
-\frac{w(k)}{2\epsilon}$ \cite{Choi:2019zpz,Choi:2019dfu,Bobev:2022eus,Bobev:2022wem}, where $w(k)={\rm Im}\,\mathcal W(k, u^*(k),\wt u^*(k))$ is the imaginary part 
of the effective twisted superpotential,
\bea
\label{twisted}
\mathcal W
&=
\sum_{i=1}^N
\left[
\frac{k}{2}
\left(\wt u_i^2-u_i^2\right)
+2\pi i\left(\wt u_i-u_i\right)
\right]
\\
&+
\sum_{i,j=1}^N\sum_{a=1}^2
\Bigg[
{\rm Li}_2\!\left(
\ii e^{\ii(\wt u_j-u_i)}
\right)
-{\rm Li}_2\!\left(
-\ii e^{\ii(\wt u_j-u_i)}
\right)
\Bigg],
\eea
evaluated at the dominant critical point $\{u^*,\wt u^*\}$, after setting $y_a=\ii$
\footnote{As in \cite{Hosseini:2026dkj} we set $y_a=\ii$ and choose the  angular ambiguities of the multivalued function $\mathcal W$ such that the long-range forces between eigenvalues
cancel and the twisted superpotential  scales
as $N^{3/2}$ at large $N$. $y_a=\ii$ can be replaced by $y_a=1$ by rescaling $z\rightarrow -\ii z$ 
}.  The general problem in this approach is to find the dominant critical point of $\mathcal W$, since the Bethe-vacuum equations $e^{\ii \partial_{u_i}\cal W} =  e^{\ii \partial_{\wt u_i}\mathcal W}=1$ are polynomial equations  in the variables $e^{\ii u_i}, e^{\ii\wt u_i}$ with multiple solutions. The dominant critical point at  large $N$ was identified in \cite{Benini:2015eyy} using matrix model techniques. The on-shell value of the twisted superpotential  is $w(k)=\frac{\pi^2}{3\sqrt{2}} k^{1/2}N^{3/2}$. The entropy of a Kerr-Newman black hole of charge $n$ is then given by $\sqrt{|w(k)|n}$ and scales  as $ k^{1/2}N^{3/2}$  for charges of order $n\sim k^{1/2}N^{3/2}$ \cite{Choi:2019zpz,Choi:2019dfu}. This identification can be extended to a general assignment of charges and  is reinforced by computations at subleading order in $N$ and their comparison with the holographic expectations \cite{Bobev:2022eus,Bobev:2022wem}.  
A set of solutions of the Bethe-vacuum equations at finite $N$,
lying on a common branch that extends down to $N=2$ and smoothly
approaches the black hole saddle for $N\gg k$, has been constructed
numerically in \cite{Hosseini:2026dkj,Hosseini:2026lvj}.
Using Bethe-vacuum data with up to $800$ significant digits,
these studies tested the perturbative large-$N$ expansions
of the on-shell twisted superpotential
and resolved exponentially
small corrections.
We conjecture that the vacuum on this branch dominates the matrix
model in the Cardy limit for \textit{all} $N$ and $k$.
Evidence for this conjecture comes from the agreement in
Figure~\ref{fig:num} between the Bethe-vacuum prediction for
$\sqrt{w(k)}$ and its numerical extraction from the large-order
behavior of the $N=2$ superconformal-index coefficients.
Assuming that this is the dominant saddle point, we find that $\sqrt{|w(k)|}= \sqrt{|{\rm Im} \mathcal W(u_i^*,\wt u_i^*)|}$ is  monotonically increasing
for all $N\ge 2$. In Figure \ref{fig:num2} we show the case $N=8$. 

At fixed $N$, $w(k)$ approaches the constant value $4 K N^2$ as $k\to\infty$. This result is confirmed numerically for $N=2,3,4$ and can be reproduced directly from the matrix model, since at $k=\infty$ only the monopole sector $m_i=\wt m_i=0$ contributes, and the integral is dominated by
$u_i=\wt u_j=\mathrm{const}\ (\mathrm{mod}\ 2\pi)$.
Near this configuration the logarithm of the integrand behaves as
$-[4KN^2-\sum_{ij}(u_i-\wt u_j)^2+\cdots]/(2\epsilon)$
\footnote{In the strict $k=\infty$ limit, a scaling argument also
predicts $C=-(N^2+1)/2$, consistently with the $N=2,3,4$ numerics.
A single Bethe vacuum gives $C=-3/4$, but its $\mathbb Z_k$ orbit,
$t^{1/k}\to e^{2\pi \ii/k}t^{1/k}$, becomes nearly degenerate at large
$k$. This effect is visible numerically for $N=2$  already at relatively small $k$.}.
For large but finite $k$, Taylor expanding the Bethe
equations on the same branch gives
$u_i,\wt u_i=\pi(N-2 i)/k+\mathcal O(k^{-2})$,
$i=1,\ldots,N$, and determines the higher orders recursively.
Stationarity of $\mathcal W$ implies
$\rd\mathcal W(k)/\rd k=\frac12\sum_{i}(\widetilde u_i^2-u_i^2)$,
which can be integrated order by order with
$\mathcal W(\infty)=4\ii KN^2$. This yields the analytic
fixed-$N$ expansion
\bea
\label{weakW}
w(k)&=4KN^2-\frac{2\pi^2N^2(N^2-1)}{3k^2}\\
&+\frac{\pi^2N^2(N^2-1)}{k^4}
\left[\frac{8N^2}{3}-\frac{2\pi^2(2N^2-3)}{45}\right]\\
&+\mathcal O(k^{-6})\,,
\eea
which approaches $4KN^2$ from below for $N\geq2$.
High-precision numerics confirm that $w(k)$ evaluated on the selected Bethe-vacuum branch at finite $k$ agrees with this expansion.

\textit{Fortuitous states and their disappearance---}%
BPS states can be divided into monotones and fortuitous states \cite{Chang:2022mjp,Choi:2023znd}. It is believed that the degeneracy of states  both at large $N$ and large charge is dominated by fortuitous states. This is certainly  true for the black hole entropy at large $N$ which cannot be explained by gravitons only. Our result supports the conjecture that fortuitous states dominate also in the Cardy limit at finite $N$, at least for large $k$. Indeed, monotones are expected to decrease with $k$, since the spectrum of gravitons for AdS$_4\times S^7/\mathbb{Z}_k$ is a $\mathbb{Z}_k$ projection of that for $k=1$. This can be checked explicitly for the sub-sector of the BPS states corresponding to the chiral ring, where the index restricts to the  Hilbert series  of $\mathrm{Sym}^N(\mathbb{C}^4/\mathbb{Z}_k)$ \cite{Cremonesi:2016nbo}. Since the total degeneracy of states grows with $k$, the number of fortuitous states is expected to increase with $k$ and eventually dominate the cohomology. Fortuitous states similarly dominate the BMN sector of $\mathcal{ N}=4$ SYM for $N=2$ at large charges, although the growth of states is only polynomial in $n$ for both fortuitous and monotone states \cite{Choi:2023znd}.  

This result is  counterintuitive. We argued in \cite{Belin:2025hsg} that the fortuitous states consist of those that can be identified  at $k=\infty$ in terms of elementary fields, with the addition of extra  operators made with monopoles. On the contrary, we see that some of the fortuitous states at $k=\infty$ disappear at finite $k$. It is conceptually possible that they are lifted by quantum corrections to the one-loop cohomology. This is known to happen for $\mathcal{ N}=4$ SYM but it affects only a small percentage of states \cite{Chang:2025mqp,Choi:2025bhi,Kim:2026rnx}. We would like to propose a more mundane explanation: as $k$ decreases, we find more potential BPS states due to monopoles but there is also room for more relations that can eliminate  fortuitous states.

We can provide an explicit example of this phenomenon by analyzing the  lowest-dimensional fortuitous states for $N=2$ as  $k$ varies.  Consider first large $k$. In \cite{Belin:2025hsg},  the monotone states were explicitly constructed by identifying the single-trace BPS operators corresponding to single gravitons at large $N$ and taking products and derivatives, modulo relations. Restricting to bosonic fields for simplicity, the single-graviton operators are $(A_iB_j)$ and $(A_iB_pA_jB_q)$ where parentheses denote traces and the $\SU(2)\times \SU(2)$ indices are symmetrized. For $N=2$, traces of more than four bifundamental fields are decomposable into products of traces with two or four. By comparing the 
full index with the index of the ring of monotones, we found  four fermionic fortuitous states at order $x^3$, transforming as $({\bf 2},{\bf 2})_0$ under the symmetry $\SU(2)\times \SU(2)\times \U(1)_t$. They are  linear combinations of a non-BPS single trace operator and the product of the non-BPS fermionic current with a graviton \cite{Belin:2025hsg,Behan:2025hbx,Behan:2026her}
\bea
    \mathcal O_f &= \epsilon^{ij}\left(\psi^A_iA_jB_kA_l+\psi^B_iB_jA_lB_k\right)\\&-\frac{3}{4}\,\epsilon^{ij}\left(B_kA_l\right)\left(\psi^A_iA_j+\psi^B_iB_j\right) ~.
\label{fort}\eea
The supersymmetry variation $\mathcal{ Q}(\mathcal O_f)$ is zero by a trace relation that holds for $N\le 2$ only.  Consider now  finite $k$. Some of the monotone states are made with monopoles. For $k>2$, the index predicts four fortuitous states that can be identified with those already found at large $k$. For $k=2$, we find instead ten fermionic fortuitous states transforming in the representation $\overline{{\bf 10}}$ of the  enhanced symmetry $\SU(4)$ commuting with the supercharge $\mathcal{ Q}$. For $k=2$ the bifundamental fields can be organized  as $Y=(A_1,A_2,B_1,B_2)$ and $\Psi=(-\psi^A_2,\psi^A_1,-\psi^B_2,\psi^B_1)$ transforming as ${\bf \overline 4}$ and ${\bf 4}$ under $\SU(4)$, respectively, and, using monopoles, we can construct gauge-invariant \textit{single-trace} operators of the form $(CC)$ and $(CCCC)$ where $C$ is either $Y$ or $\Psi$.  The four fortuitous states $\mathcal O_f$ found at large $k$ can be  completed to an $\SU(4)$ multiplet  of ten fortuitous states of   schematic form $(\Psi YYY)+(YY)(\Psi Y)$. Consider finally  $k=1$. We find no fortuitous states at order $x^3$. We give an intuitive reason here and refer to Supplemental Material for more details. At $k=1$, $Y$ and $\Psi$ are themselves gauge invariant when combined with monopoles. The index factorizes into a product 
$\mathcal{ I}(x)=\mathcal{ I}_\text{red}(x)\times \mathcal{ I}_\text{free}(x)$ of the index of four free chiral multiplets and a \textit{reduced} index. $\mathcal{ I}_\text{free}$ is fully reproduced by considering products of $(Y)$ and $(\Psi)$ and their derivatives. The reduced index $\mathcal{ I}_\text{red}$ is reproduced  by taking products of \textit{single-trace} operators of schematic form $(YY)$, $(\Psi Y)_\text{traceless}$, $(\Psi\Psi+YDY)$ and $(\Psi DY -Y D\Psi)$ and their derivatives modulo relations 
\footnote{This is the $S_2$ graviton multiplet discussed in Appendix B.2 of \cite{Belin:2025hsg} (see Supplemental Material).}. In particular,  the operators $(CCCC)$ are not independent and they are decomposable into products of $(CC)$ and $(C)$. The bifundamental fields $C$ behave effectively as two-by-two matrices. 
Even gauge-invariant operators made with fundamental fields, like $(A_iB_pA_jB_q)$, which are non-decomposable and independent for all $k\ge 2$, can be written as products of  monopole operators, revealing new relations at $k=1$.   In particular, the single-trace part of the fortuitous states \eqref{fort} is decomposable and $\mathcal{ O}_f$ can be written as a product of gravitons \footnote{The relevant identity is (4.49) in \cite{Choi:2026faq}
(see Supplemental Material).}. The  reduced cohomology can be understood only in terms of monotones up to order $x^{5}$ and we believe that the first fortuitous state appears at much higher order, as in $\mathcal{ N}=4$ SYM \cite{Choi:2023znd}.

The case of $N=1$ is even more extreme, and all gauge invariants are BPS since there are no trace relations and the $\mathcal{ Q}$-action vanishes. However, not all of them are monotones. For example, the $\U(1)$ under which all the fields carry the same charge is a symmetry for $N=1$ but not for $N>1$ and  the corresponding fermionic current $\epsilon_{ij}\left(\psi^A_iA_j+\psi^B_iB_j\right)$ is a fortuitous state for $k>1$. Once again, for  $k=1$ the BPS states are freely generated by $(Y)$ and $(\Psi)$ and their derivatives and the fermionic current decomposes into a product of gravitons. For  $k=1$ all the BPS states are  monotones \cite{Belin:2025hsg}.  

We see that monopoles enlarge the number of operators but also introduce new relations.
We conjecture that this mechanism is general and responsible for the $\sqrt{k}N^{3/2}$ scaling of the degeneracies at large $N$, and leave a proof for future work.

\textit{Acknowledgments---}%
We would like to thank Nikolay Bobev, Chi-Ming Chang, Seok Kim and Juan Maldacena for useful discussions. 
AB, PS, AZ and RV are partially supported by the INFN. SMH was supported by UKRI under the UK government's Horizon Europe funding guarantee (Grant No.~EP/Y027604/1) and STFC under Grant No.~ST/X000494/1.

\bibliographystyle{apsrev4-2}   
\bibliography{puzzle}

\clearpage
\onecolumngrid 

\begin{center}
  \large\bfseries Supplemental Material
\end{center}

\setcounter{equation}{0}
\renewcommand{\theequation}{S\arabic{equation}}

\setcounter{figure}{0}
\renewcommand{\thefigure}{S\arabic{figure}}

\setcounter{table}{0}
\renewcommand{\thetable}{S\arabic{table}}

\section*{\boldmath The BPS cohomology for $N=2$ and $k=1$}
In this appendix we fill a gap in \cite{Belin:2025hsg} and discuss the ABJM BPS cohomology for $N=2$ and $k=1$ up to order $x^4$. We refer to \cite{Belin:2025hsg} for conventions and details about supersymmetry transformations and the role of monopole operators. We organize the four bosonic  fields $(A_1,A_2,B_1,B_2)$ in a multiplet $Y$ transforming in  the $\overline{\mathbf 4}$ of $\SU(4)$ with weights $x^{1/2}y_a$, $a=1,2,3,4$, and  the fermions in a multiplet $\Psi=(-\psi_2^A,\psi_1^A,-\psi_2^B,\psi_1^B)$ transforming in  the $\mathbf 4$ with weights $x^{3/2}y_a^{-1}$. At $k=1$, the $Y$ and $\Psi$ can be made gauge invariant by combining them with the monopole operators $T^{(\pm 1,0)}$. All monopoles in this appendix have ${\bf m}=\wt{{\bf m}}$ and we abbreviate $T^{({\bf m},\wt{{\bf m}})}$ as $T^{\bf m}$. The free part of the cohomology is freely generated by the corresponding operators  $(Y)$ and $(\Psi)$, and the index reads
\begin{equation}
\mathcal I_\text{free}(x;y)
=
\prod_{n=0}^{\infty}\prod_{a=1}^{4}
\frac{1-x^{3/2+2n}y_a^{-1}}
     {1-x^{1/2+2n}y_a}.
\end{equation}
The index of the reduced cohomology is $\mathcal I_{\mathrm{red}}(x;1)
=1+10x+20x^2+20x^3+65x^4+O(x^5)$ fully refined as
\begin{equation}
1+\chi_{\overline{\mathbf{10}}}(q)x
+\left(
\chi_{\overline{\mathbf{35}}}(q)
-\chi_{\mathbf{15}}(q)
\right)x^2+\left(
\chi_{\overline{\mathbf{84}^{\prime\prime}}}(q)
-\chi_{\overline{\mathbf{70}}}(q)
+\chi_{\mathbf 6}(q)
\right)x^3+\left(
\chi_{\overline{\mathbf{165}}}(q)
+2\chi_{\overline{\mathbf{45}}}(q)
-\chi_{\overline{\mathbf{189}}}(q)
-1
\right)x^4
+O(x^5),\nonumber
\end{equation}
where $\chi_R(q)$ is the $\SU(4)$ character of the representation $R$ and $q_a=y_a^{-1}$ are the weights of the fundamental representation, $\chi_{\bf 4}(q)=\sum_{a=1}^4 q_a=\sum_{a=1}^4 y_a^{-1}$.
The reduced monotone cohomology is generated by the $S_2$ graviton multiplet  $\{(YY),(\Psi Y)_{\mathrm{traceless}}, (\Psi\Psi+YDY), (\Psi DY-Y D\Psi)\}$ with respective weights $\{\chi_{\overline{\mathbf{10}}} (q) x\, , \chi_\mathbf{15}(q)  x^2\, ,\chi_{\mathbf 6}(q)  x^3\, , \chi_{\mathbf 1}(q)  x^4
\}$. $D$ indicates the derivative and, above and below,  the multiplets are given in a schematic form, rather than with exact relative coefficients. The multiplet $(YY)$ is made with $\{(A_iB_j),T^{(1,1)}\det B, T^{(-1,-1)}\det A\}$ corresponding to the decomposition $\overline{\mathbf{10}}\rightarrow (2,2)_0+(1,3)_2+(3,1)_{-2}$ under $\SU(2)\times \SU(2)\times \U(1)_t$. Similarly,  $(\Psi Y)_{\mathrm{traceless}}$ contains the $\SU(4)$ fermionic currents and it is made of $\{(\psi_{(i}^A A_{j)}), (\psi^B_{(i} B_{j)}), \epsilon^{ij}(\psi^A_iA_j-\psi^B_iB_j), (T^{(-1,-1)}\det \psi^B A), (T^{(1,1)}\det \psi^A B)\}$ corresponding to the decomposition $\overline{\mathbf{15}}\rightarrow (3,1)_0+(1,3)_0+(1,1)_0+(2,2)_{-2}+(2,2)_{2}$. Here $\det \psi^A B$ stands for $\epsilon_{ab}\epsilon_{\bar a \bar b}\psi^A_{a\bar a}B_{b\bar b}$. The operators $T^{\pm(1,1)}$ carry only $\U(1)\times\U(1)$ electric charge and can be made gauge invariant by dressing them with operators that are invariant under the gauge group $\SU(2)\times \SU(2)$ and carry opposite $\U(1)\times\U(1)$ charge. Due to monopoles, it is difficult to properly take into account the relations between operators. We propose that the multi-gravitons up to order $x^4$ are
$$
\begin{array}{|c|l|l|}
\hline
\text{degree}
&
\text{schematic operators}
&
\SU(4)
\\
\hline\hline
x
&
(YY)
&
\overline{\mathbf{10}}
\\
\hline
x^2
&
(YY)^2,\quad (Y\Psi)
&
\overline{\mathbf{35}}-\mathbf{15}
\\
\hline
x^3
&
(YY)^3,\quad (\Psi\Psi+YDY),\quad (YY)(Y\Psi),\quad D(YY)
&
\overline{\mathbf{84}^{\prime\prime}}+\mathbf 6-\overline{\mathbf{70}}
-\overline{\mathbf{10}}+\overline{\mathbf{10}}
\\
\hline
x^4
&
(YY)^4,\quad (YY)^2(Y\Psi),\quad
(Y\Psi)^2,\quad
(YY)(\Psi\Psi+Y DY),\quad
(\Psi DY-YD\Psi)
&
\overline{\mathbf{165}}
-\overline{\mathbf{189}}
-\overline{\mathbf{35}}
+\mathbf{15}
+\overline{\mathbf{45}}
-\mathbf 1
\\
&
(YY)D(YY),\quad D(Y\Psi)
&
\overline{\mathbf{45}}+\overline{\mathbf{35}}-\mathbf{15}
\\
\hline
\end{array}
$$
The table has been constructed assuming that traces of more than two operators are decomposable, as discussed in the main text, and the bosonic letters $Y$ are fully symmetrized. The latter condition is the $\SU(4)$ completion of the separate symmetrization of $A_i$ and $B_i$ 
implied by the supersymmetric algebra. For example, the 55 operators $(YY)^2$ are reduced by trace relations to 35 operators corresponding to the representation $\overline{\mathbf{35}}=\mathrm{Sym}^4 \overline{\mathbf{4}}$. More generally, the independent BPS chiral operators of the form $(YY)^n$  are those transforming in the $\mathrm{Sym}^{2n} \overline{\mathbf{4}}$ representation. This correctly reproduces the Hilbert series of the reduced chiral ring. It is indeed shown in \cite{Cremonesi:2016nbo} that the Hilbert series coincides with that of the symmetric product of $\mathbb{C}^4$, as appropriate for a theory of M2-branes in flat space,  and decomposes as  $\text{HS}_{\mathrm{Sym}^2(\mathbb{C}^4)}(x)=\text{HS}_\text{red}(x) \text{HS}_{\mathbb{C}^4}(x)$ where
\begin{equation}
\text{HS}_\text{red}(x)=\frac{1+6 x+x^2}{(1-x)^4}=1+10 x+35 x^2+84 x^3+165 x^4+O(x^5)\, .\end{equation}
Notice that the same chiral ring can be reproduced by assuming that $Y$ are  four mutually commuting adjoint matrices of $\U(2)$, where the $\U(1)$ part factorizes. Effectively, in the reduced cohomology $Y$ behave as four commuting $\SU(2)$ adjoints. Similarly, $(YY)(\Psi Y)$ contributes only $\overline{\mathbf{70}}+\overline{\mathbf{10}}$, where  the three $Y$ have been symmetrized. This statement is reinforced by the fact that $(YY)(\Psi Y)$ can be obtained by applying to $(YY)^2$ the supercharges that commute with $\mathcal{ Q}$ and transform as $\mathbf{6}$. Similar arguments have been used to identify the remaining BPS operators. The multi-graviton table reproduces the index up to order $x^4$. We see that at this order there is no evidence for the existence of fortuitous states, although they could in principle exist and cancel out in the index. A similar analysis at order $x^5$ still predicts no fortuity. At higher orders, descendant operators obtained by taking derivatives and their relations  start playing a role and the analysis is more complicated. A rudimentary toy model where the generators of the $S_2$ graviton multiplet are realized in terms of four bosons $c_i$ and four fermions $f_i$ corresponding to the eigenvalues of $Y$ and $\Psi$ considered as $\SU(2)$ adjoint matrices reproduces the index in terms of monotones  up to order $x^{12}$. It would be interesting to see whether the first fortuitous state appears at order $O(x^{12})$ as in $\mathcal{ N}=4$ SYM \cite{Choi:2023znd}.

In particular, the four fortuitous states \eqref{fort} found at large $k$ should be expressible in terms of multi-gravitons. The relevant relation was found  in \cite{Choi:2026faq} and states that, up to gravitons and $\mathcal{ Q}$-exact terms, \begin{equation} \mathcal{ O}_f \sim \det A \det \psi^A B + \det B \det \psi^B A \Big |_{(2,2)}\label{kim}\end{equation}
where, on the right-hand side, the product of operators is restricted to the representation $(2,2)$ of the global symmetry $\SU(2)\times\SU(2)$. Reference \cite{Choi:2026faq} suggested reconsidering the notion of fortuity in ABJM at large $k$ based on identities of this kind. We propose here that this identity is indeed relevant for the disappearance of fortuitous states but at $k=1$. At large $k$, \eqref{kim} expresses $\mathcal O_f$ as a product of non-gauge-invariant determinant operators.  Determinants can be made gauge invariant by multiplying them with monopole operators. For the determinants $\det A , \det B ,\det \psi^A B , \det \psi^B A$ this can happen only at $k=1$, where, combined with $T^{\pm(1,1)}$, they become part of the multiplets $(YY)$ and $(\Psi Y)$. We propose that for $k=1$, \eqref{kim} becomes an identity relating $\mathcal O_f$ to the product of gauge-invariant dressed monopole operators, $\mathcal O_f\sim (YY)(Y\Psi)\big |_{(2,2)}$, thus decomposing $\mathcal O_f$ as a linear combination of  products of gravitons.

\end{document}